%%%%%%%%%%%%%%%%%%%%%%%%%%%%%%%%%%%%%%%%%%%%%%%%%%%%%%%%%%%%%%%%%%%%%%%%%%%%
\input harvmac

\def\pl#1#2#3{Phys. Lett. {\bf #1B} (#2) #3}

\def\ap#1#2#3{Ann. Phys. {\bf #1} (#2) #3}

\def\cmp#1#2#3{Comm. Math. Phys. {\bf #1} (#2) #3}

\def\ap#1#2#3{Ann.~Phys. {\bf #1} (#2) #3}

%%%%%%%%%%%%%%%  Rublenye bukvy   %%%%%%%%%%%%%%%%%
\def\IB{\relax\hbox{$\inbar\kern-.3em{\rm B}$}}
\def\IC{\relax\hbox{$\inbar\kern-.3em{\rm C}$}}
\def\ID{\relax\hbox{$\inbar\kern-.3em{\rm D}$}}
\def\IE{\relax\hbox{$\inbar\kern-.3em{\rm E}$}}
\def\IF{\relax\hbox{$\inbar\kern-.3em{\rm F}$}}
\def\IG{\relax\hbox{$\inbar\kern-.3em{\rm G}$}}
\def\IGa{\relax\hbox{${\rm I}\kern-.18em\Gamma$}}
\def\IH{\relax{\rm I\kern-.18em H}}
\def\IK{\relax{\rm I\kern-.18em K}}
\def\IL{\relax{\rm I\kern-.18em L}}
\def\IP{\relax{\rm I\kern-.18em P}}
\def\IR{\relax{\rm I\kern-.18em R}}
\def\IZ{\relax\ifmmode\mathchoice{
\hbox{\cmss Z\kern-.4em Z}}{\hbox{\cmss Z\kern-.4em Z}}
{\lower.9pt\hbox{\cmsss Z\kern-.4em Z}}
{\lower1.2pt\hbox{\cmsss Z\kern-.4em Z}}
\else{\cmss Z\kern-.4em Z}\fi}
\def\II{\relax{\rm I\kern-.18em I}}

\def\ndt{{\noindent}}
\def\bx{{\bf G}}

%%%%%%%% Calligraphic letters  %%%%%%%%%%%%%

\def\CD{{\cal D}}
\def\CE{{\cal E}}
\def\CF{{\cal F}}

\def\CI{{\cal I}}

\def\CM{{\cal M}}

\def\CO{{\cal O}}
\def\CP{{\cal P}}

\def\CS{{\cal S}}

\def\CU{{\cal U}}

\def\CZ{{\cal Z}}

%%%%%%%%%%%% Derivatives  %%%%%%%%%%%
\def\p{\partial}
\def\pb{\bar{\partial}}

\def\dd{{\rm d}}
%%%%%%%%%%% letters with bar %%%%%%%%

\def\sb{\bar{s}}
\def\tb{\bar{t}}
\def\zb{\bar{z}}

%%%%%%%%%% Math symbols %%%%%%%%%%%%%

\def\Det{{\rm Det}}

%%%%%%%%%%%%%% Lie algebras %%%%%%%%%%%%%%%%%%%%%%

\def\inbar{\,\vrule height1.5ex width.4pt depth0pt}

\font\cmss=cmss10 \font\cmsss=cmss10 at 7pt

%%%%%%%%%%%% Greek %%%%%%%%%%%%
\def\a{{\alpha}}
\def\b{{\beta}}

\def\z{{\zeta}}

\def\vf{{\varphi}}
\def\m{{\mu}}

\def\l{{\lambda}}
\def\s{{\sigma}}
\def\t{{\theta}}

%%%%%%%%%%%%%%%%%%%%%%%%%%%
% REFS
\lref\burns{D.~Burns, in, ``Twistors and Harmonic Maps'',
Lecture. Amer. Math. Soc. Conference, Charlotte, NC, 1986}
\lref\cg{E.~Corrigan, P.~Goddard, ``Construction of instanton
and monopole solutions and reciprocity'', \ap{154}{1984}{253}}
\lref\donaldson{S.K.~Donaldson, ``Instantons and Geometric Invariant Theory",
Commun. Math. Phys. 93 (1984) 453-460.}
\lref\nikthes{N.~Nekrasov, ``Topological theories and Zonal Spherical
Functions'', ITEP publications, 1995 (in Russian)}
\lref\nakajima{H.~Nakajima, ``Lectures on Hilbert Schemes
of Points on Surfaces''\semi
AMS University Lecture Series, 1999, ISBN 0-8218-1956-9. }
%{\tt http://www.kusm.kyoto-u.ac.jp/~nakajima/TeX.html}}
\lref\neksch{N.~Nekrasov, A.~S.~Schwarz,
{\tt hep-th/9802068}, \cmp{198}{1998}{689}}
\lref\freck{A.~Losev, N.~Nekrasov, S.~Shatashvili, ``The Freckled Instantons'',
{\tt hep-th/9908204}, Y.~Golfand Memorial Volume, M.~Shifman Eds.,
World Scientific, Singapore, to appear}
\lref\rkh{N.J.~Hitchin, A.~Karlhede, U.~Lindstrom, and M.~Rocek,
Comm.~Math.~Phys. {\bf 108} (1987) 535}
\lref\wilson{G.~ Wilson, ``Collisions of Calogero-Moser particles and
adelic Grassmannian", Invent. Math. 133 (1998) 1-41.}
\lref\abs{O.~Aharony,
M.~Berkooz, N.~Seiberg, hep-th/9712117,
Adv.Theor.Math.Phys. 2 (1998) 119-153}
\lref\abkss{O.~Aharony, M.~Berkooz, S.~Kachru, N.~Seiberg, E.~Silverstein,
hep-th/9707079,
Adv.Theor.Math.Phys. 1 (1998) 148-157}
\lref\witsei{E.~Witten, N.~Seiberg, hep-th/9908142, JHEP 9909 (1999) 032}
\lref\nikdui{N.~Nekrasov,
``On a duality in Calogero-Moser-Sutherland systems'',
hep-th/9707111}
\lref\dual{V.~Fock, A.~Gorsky, N.~Nekrasov, V.~Rubtsov, ``Duality
in Integrable Systems and Gauge Theories'', hep-th/9906235}
\lref\ruij{S.~M.~Ruijsenaars, Comm. Math. Phys. {\bf 110} (1987) 191-213\semi
S.~M.~Ruijsenaars and H.~Schneider, Ann. Phys. (NY) 170 (1986) 370}
\lref\relcal{A.~Gorsky, N.~Nekrasov, hep-th/9401017, Nucl. Phys. {\bf B}436 (1995) 582}
\lref\kkk{D. Kazhdan, B. Kostant, S.~Sternberg, ``Hamiltonian Group Actions
and Dynamical Systems of Calogero Type", Commun. Pure and Appl. Math.
31 (1978) 481-507.}
\lref\GKMMM{A.~Gorsky, I.~Krichever, A.~Marshakov, A.~Mironov and A.~Morozov,
``Integrability and Seiberg-Witten Exact Solution", Phys.Lett.,  B355 (1995)
466-474, hep-th/9505035}
\lref\GMM{A.~Gorsky, S.~Gukov, A.~Mironov,
``Multiscale N=2 SUSY field theories, integrable systems and their
 stringy/brane origin -- I",
Nucl.Phys. B517 (1998) 409-461, hep-th/9707120 }
\lref\GMMM{A.~Gorsky, A.~Marshakov, A.~Mironov, A.~Morozov,
Nucl.Phys., ``RG Equations from Whitham Hierarchy", B527 (1998)
690-716; hep-th/9802007} \lref\MW{E.~Martinec and N.~Warner,
``Integrable systems and supersymmetric gauge theory", Nucl.Phys.,
{ 459} (1996) 97-112, hep-th/9509161} \lref\DW{ R.~Donagi and
E.~Witten, ``Supersymmetric Yang-Mills Systems And Integrable
Systems", Nucl.Phys., {\bf B460} (1996) 299-344, hep-th/9510101}
\lref\BMMM{H.W.~Braden, A.~Marshakov, A.~Mironov and A.~Morozov,
``The Ruijsenaars-Schneider Model in the Context of Seiberg-Witten
Theory", hep-th/9902205 } \lref\BK{See for example contributions
in, ``Integrability: the Seiberg-Witten and Whitham equations",
eds H.W. Braden and I.M. Krichever, Amsterdam: Gordon and Breach
Science Publishers.} \lref\cmb{See  the contributions of G.Wilson,
I.Krichever, N.Nekrasov and H.W. Braden in ``Proceedings of the
Workshop on Calogero-Moser-Sutherland models" CRM Series in
Mathematical Physics, Springer-Verlag.} \lref\sepa{A.~Gorsky,
N.~Nekrasov, V.~Rubtsov, ``Hilbert Schemes, Separated Variables,
and D-branes'', hep-th/9901089} \lref\mrdone{M.~Douglas, ``Branes
within Branes'', hep-th/9512077} \lref\mrdtwo{M.~Douglas, ``Gauge
Fields and D-branes'',  hep-th/9604198, J.~Geom.~Phys. 28 (1998)
255-262} \lref\mmms{S.~Terashima, ``Instantons in the $U(1)$
Born-Infeld Theory and Noncommutative Gauge Theory'',
hep-th/9911245, \pl{477}{2000}{292-298}\semi M.~Mari{\~n}o,
R.~Minasian, G.~Moore, A.~Strominger, ``Non-linear Instantons from
Supersymmetric p-Branes'', hep-th/9911206}
\lref\laz{C.~Lazaroiu,
``A noncommutative-geometric interpretation of the resolution of
equivariant instanton moduli spaces'', hep-th/9805132}

\Title{\vbox {\baselineskip 10pt \hbox{PUPT-1900}
\hbox{ITEP-TH-51/99} \hbox{NSF-ITP-99-144} \hbox{hep-th/9912019}
{\hbox{   }}}} {\vbox{\vskip -30 true pt \centerline{SPACE-TIME
FOAM FROM }
\medskip
\centerline{NON--COMMUTATIVE INSTANTONS}
\medskip
\vskip4pt }}
\vskip -20 true pt
\centerline{ Harry W.~Braden$^{1}$, Nikita A.~Nekrasov$^{2}$}
\smallskip\smallskip
\centerline{$^{1}$ \it Dept. of Mathematics and Statistics,
 University of Edinburgh, Edinburgh EH9 3JZ, Scotland}
\centerline{$^{2}$ \it Joseph Henry
Laboratories, Princeton University, Princeton, New Jersey 08544}
\centerline{$^{1,2}$ \it Institute for Theoretical and Experimental
Physics, 117259 Moscow, Russia}
\medskip
\centerline{e-mail: \sl hwb@ed.ac.uk, nikita@feynman.princeton.edu}
\vskip 2cm
\centerline{\bf ABSTRACT}
\vskip 1cm
\ndt We show that a  $U(1)$ instanton on non-commutative
${\IR}^4$ corresponds to a non-singular $U(1)$ gauge field
on a commutative K\"ahler manifold $X$ which is a blowup
of ${\IC}^2$ at a finite number of points. This gauge field on $X$
obeys Maxwell's equations in addition to the susy
constraint $F^{0,2} =0$.
For instanton charge $k$ the manifold
$X$ can be viewed as a space-time foam
with $b_{2} \sim k$. A direct connection with integrable systems of
Calogero-Moser type is established.
We also make some comments on the non-abelian case.

\Date{12/99}
\vfill\eject
\newsec{Introduction}

The moduli space ${\CM}_{k,N}$ of the charge $k$
instantons in the gauge group $U(N)$  shows up in many
problems in mathematical physics and more recently in
string theory.  This space is non-compact, due to the well-known
phenomenon of instantons being able to shrink, and there are
several celebrated ways of (partially) compactifying this space.
One option, motivated by  the Uhlenbeck's theorem concerning the
extension of finite action gauge fields to an isolated
point, is to add to the space ${\CM}_{k,N}$ the space
${\CM}_{k-1,N} \times X$ which corresponds to a single
point-like instanton in the background of the smooth
charge $k-1$ instanton. One then further adds the space
${\CM}_{k-2,N} \times {\rm Sym}^2 X$ corresponding to the
pairs of the point-like instantons, and so on. In this way one
obtains the Donaldson compactification:
\eqn\dona{{\overline\CM}_{k,N}^{D} = {\CM}_{k,N} \cup
{\CM}_{k-1,N} \times X \cup {\CM}_{k-2,N} \times {\rm Sym}^2 X
\ldots \cup {\rm Sym}^{k}X.}
If the space-time $X$ is a projective surface $S$ with the K\"ahler form
$\omega$ then there
is a finer compactification, the space of the torsion
free sheaves. This compactification ${\overline\CM}^{G}_{k,N}$
is the space of all $\omega$-stable torsion free sheaves of rank $N$ and
second Chern class $c_{2} = k$. In the case $S = {\IC\IP}^2$
one can study the sheaves which are trivial when restricted to
the projective line ${\IC\IP}^1$ at infinity. This
space ${\overline M}_{k,N}$ has an ADHM-like description. It was shown in
\neksch\ that this space parameterises instantons on
the non-commutative space ${\IR}^4$ where the degree of the
non-commutativity is related
to the metric on the space ${\overline M}_{k,N}$.
This deformation of the ADHM equations also arises in the study of integrable
systems of Calogero-Moser type \nikthes\wilson\cmb ; these same models have
appeared in connection with supersymmetric gauge theories
\GKMMM\MW\DW\GMM\BMMM\BK\sepa\
and admit a brane description \abkss\abs\witsei.

An outline of our paper is as follows. In section two
we review the physical motivation for our work. Next we will review
the deformed ADHM equations we are interested in, paralleling the usual
ADHM construction. For a particular choice of complex structure we find
the resulting equations describe appropriate holomorphic data. Our aim is to
show this actually describes holomorphic bundles on a ``blown up"
spacetime. We begin  (section 4) by focusing attention on the abelian
setting which is rather illustrative, and an unexpected richness is found
for sufficiently large charge. A direct correspondence with the
Calogero-Moser integrable system is established.
Section 5 continues with the nonabelian situation.
We conclude with a brief discussion.

\newsec{Physical motivation}

Consider the theory on a stack of $N$ D3
branes in the Type IIB string theory. Add a collection of
$k$ D-instantons and  switch on a constant, self-dual $B$-field
along the D3-brane worldvolume. The D-instantons cannot escape
the D3-branes without breaking supersymmetry \witsei.
From the point of view of the gauge theory living on the D3-branes,
the D-instantons are represented by field configurations with
non-trivial instanton charge \mrdone.
Those instantons which shrink to zero size become D-instantons, and such can
escape from the D3-brane worldvolume. Therefore, in the
presence of the $B$-field, one cannot make the instanton shrink.
One realization of this scenario was suggested in \neksch\
where it was proposed to view the D-instantons within the D3-brane
with $B$-field as the instantons of a gauge theory on a
non-commutative space-time.
However, the non-commutative gauge theory as arising in the zero slope
limit of the open string theory in a particular regularization can be mapped
to the ordinary commutative gauge theory, as shown in \witsei.
Therefore one is led to the following
puzzle in the $N=1$ case: how is it possible
for the $U(1)$ gauge field on ${\IR}^4$ to have a non-trivial
instanton charge? It is easy to
show that a non-trivial charge is incompatible with the vanishing
of $F$ at infinity.

At the same time, one can look at what is happening from the point of view
of the D-instantons. Equally, by T-duality one can study the D0-D4 system,
and look at the quantum mechanics of D0-branes. The latter has
a low-energy target space which coincides with the resolution of the
singularities ${\overline M}_{k,N}$ of the instanton moduli space.

One can imagine probing the instanton gauge field as in \mrdtwo\
(perhaps employing further T-dualities).
When the $B$-field is turned on the probed gauge field is given by
the deformed ADHM construction described below. As we shall see, the
resulting gauge fields are singular unless one changes the topology of
the space-time.

We suggest that this is what indeed happens. In this way we resolve
the paradox with the $U(1)$ gauge fields, since if the space-time
contains non-contractible two-spheres (and this is precisely what
we shall get) then the $U(1)$ gauge field can have a non-trivial instanton
charge.
As far as the concrete mechanism for such a topology change within
string theory is concerned this will be left to future work.

\newsec{The deformed ADHM construction}

From now on we make the change of notation: $k = v, N = w$.
Let $V$ and $W$ be hermitian complex vector spaces of
dimensions $v$ and $w$ respectively. Let
$B_{1}$ and $B_{2}$ be the maps from $V$ to itself,
$I$ be the map from $W$ to $V$ and finally
let $J$ be the map from $V$ to $W$.
We can form a sequence of linear maps
\eqn\monad{V \longrightarrow^{\kern -10pt \s \quad}
V \otimes {\IC}^2 \oplus W \longrightarrow^{\kern -10pt \tau \quad } V}
where
\eqn\mps{{\s} = \pmatrix{-B_{2}\cr B_{1} \cr J}, \qquad {\tau} =
\pmatrix{ B_{1} & B_{2} & I}.}
We will also use
$$ {\s_z} = \pmatrix{-B_{2} + z_2\cr B_{1}- z_1 \cr J}, \qquad {\tau}_z =
\pmatrix{ B_{1}  - z_1 & B_{2} - z_2 & I}.
$$

Suppose now that the matrices $(B_{1,2}, I, J)$
obey the following equations:
\eqn\mmnts{\eqalign{\tau \s \quad & = \quad {\zeta}_{c} {\bf 1}_{V}, \cr
\tau \tau^{\dagger} \quad & = \quad \Delta + \zeta_{r} {\bf 1}_{V}, \cr
{\s}^{\dagger}{\s} \quad & =  \quad \Delta - \zeta_{r} {\bf 1}_{V}. \cr}}
Let us collect
the numbers $({\z}_r, {\rm Re} {\z}_{c}, {\rm Im} {\z}_{c})$ into a
three-vector ${\vec{\z}} \in {\IR}^3$.
When ${\vec{\z}}=0$ these equations, together with the injectivity and
surjectivity of ${\s_z}$ and ${\tau}_z$ respectively, yield the standard
ADHM construction. If one relaxes the injectivity condition
then one gets the Donaldson compactification of
the instanton moduli space. In the nomenclature of Corrigan and Goddard \cg \
describing charge $v$ $SU(w)$ instantons,
$${{\bf \Delta}= \pmatrix{-B_{2}& B_{1}\sp\dagger\cr B_{1} &
       B_{2}\sp\dagger\cr J&I\sp\dagger}},$$
and ${\bf \Delta}\sp\dagger{\bf \Delta}=\Delta\otimes {\bf 1}_2$ corresponds
to the equations \mmnts\  when ${\vec{\z}}=0$.
We are considering a deformation of the standard ADHM equations.

The space of all matrices
$(B_{1}, B_{2}, I, J)$ is  a hyperk\"ahler vector space
and the equations \mmnts\ may be interpreted as $U(k)$ hyperk\"ahler moment
maps \rkh.  In particular by performing an $SU(2)$ transformation
\eqn\suro{ \pmatrix{B_{1} \cr B_{2}\cr} \mapsto \pmatrix{ {\a} B_{1} - {\b}
B_{2}^{\dagger} \cr {\bar\a} B_{2} + {\bar \b} B_{1}^{\dagger} \cr},
\pmatrix{I \cr J\cr}
\mapsto \pmatrix{ {\a} I - {\b} J^{\dagger} \cr
{\bar\a} J + {\bar \b} I^{\dagger} \cr},}
with $\vert \a \vert^2 + \vert \b \vert^2 =1$, we can always rotate
$\vec\zeta$ into a vector $({\z}_r , 0, 0)$. Such a transformation corresponds
to singling out a particular complex structure on our data, for which
$z=(z_1,z_2)$ are the holomorphic coordinates on the Euclidean
space-time. Further we may choose the complex structure such that
${\z}_r>0$.

The  moduli space $\overline{M}_{v,w}$ is the space of
solutions to the equations \mmnts\ up to a symmetry transformation
\eqn\scndr{\left( B_{1}, B_{2}, I, J \right)
\mapsto \left( g^{-1}B_{1}g, \, g^{-1} B_{2} g, \, g^{-1} I,  \,  J g  \right)}
for $g  \in U(w)$.
It is the space of freckled instantons on ${\IR}^4$ in the sense  of \freck,
a ``freckle" simply being a point at which $\sigma_z$ fails to be
injective.\foot{Of course the algebra of functions at such points
carries interesting information: it is a finite-dimensional commutative
associative algebra which still may have nilpotents. In this sense the
freckle is a ``fat point" (or ``zero dimensional subscheme").}
Observe that for ${\z}_r>0$ \mmnts\ shows that
${\tau}_{z}{\tau}_{z}^{\dagger}$ is invertible and
$\tau_z$ is surjective.

{}One can learn from \nakajima\ that the deformed ADHM data
parameterise the (semistable) torsion free sheaves on ${\IC\IP}^2$
whose restriction on the projective line ${\ell}_{\infty}$
at infinity is trivial.
Each torsion free sheaf ${\CE}$ is included into the exact
sequence of sheaves
\eqn\exct{0 \longrightarrow {\CE} \longrightarrow
{\CF} \longrightarrow {\CS}_{Z} \longrightarrow 0}
where ${\CF}$ is a holomorphic bundle ${\CE}^{**}$ and ${\CS}_Z$ is a
skyscraper sheaf supported at points, the set $Z$ of freckles\freck.
{}From this exact sequence one learns that
\eqn\chrns{{\rm ch}_{i}({\CE}) = {\rm ch}_{i}({\CF}) - \# Z \delta_{i,2}.}

\subsec{Constructing the gauge field}

The fundamental object is the solution of
\eqn\ferm{{\CD}^{\dagger}_{z} \Psi_{z} = 0, \qquad {\Psi}_{z} : W \to
V \otimes {\IC}^2 \oplus W}
where $${\CD}^{\dagger}_{z} = \pmatrix{{\tau}_{z} \cr {\s}^{\dagger}_{z} }.$$
We shall need the components
\eqn\cmps{\Psi_{z} = \pmatrix{\Psi_{1} \cr \Psi_{2} \cr \chi}
= \pmatrix{\vf \cr \chi}, \quad
\Psi_{1,2} \in V,\,\, {\vf} \in V \otimes{\IC}^2 ,\,\,
{\chi} \in W.}
The solution of \ferm\
is not uniquely defined and one is free
to perform a $GL(w, {\IC})$ gauge transformation, $$\Psi_{z} \to \Psi_{z}
\, g(z,
{\zb}), \quad
g(z, {\zb}) \in {\rm GL}(w, {\IC}). $$
This gauge freedom can be partially fixed
by normalising the vector $\Psi_{z}$ as follows:
\eqn\nrm{\Psi_{z}^{\dagger} \Psi_{z} = {\bf 1}_{W}.}
With this normalisation the $U(w)$ gauge field is given by
\eqn\gf{A = {\Psi}^{\dagger}_{z} {\rm d} {\Psi}_{z},}
and its curvature is given by
\eqn\crvt{F = {\Psi}^{\dagger}_{z} {\rm d}{\CD}_{z} {1\over{
{\CD}^{\dagger}_{z} {\CD}_{z}}}
{\rm d}{\CD}^{\dagger}_{z} {\Psi}_{z}.}

More explicitly, $${\CD}^{\dagger}_z {\CD}_z = \Delta_z \otimes {\bf 1} +
{\bf 1}_{V} \otimes \zeta^{a} {\s}_{a},$$ hence
$$
{1\over {\CD}^{\dagger}_z {\CD}}_z = {1\over{{\Delta}^2_z - \vec\zeta^2}}
\left( \Delta_z \otimes {\bf 1} - {\bf 1}_{V} \otimes {\zeta}^{a}{\s}_{a}
\right).
$$
Formula \crvt\ makes sense for $z \in X^{\circ}\equiv{\IR}^4 \setminus Z$,
where $X^{\circ}$ is the complement in ${\IR}^4$ to the set $Z$ of points
(freckles) at which
\eqn\discr{{\rm Det} \left( \Delta^2_z -  \vec\zeta^2 \right)= 0. }
Now it is a straightforward exercise to show that on $X^{\circ}$
\eqn\sdp{F^{+} ={1\over2}\left(F+\phantom{.}\sp\star F\right)=
\, {\vf}^{\dagger} {1\over {{\Delta}^2_{z} - {\vec \zeta}^2}} {\vf}
\,  \hat \zeta }
where
$\hat\z =
\z_{r} {\varpi}_{r} + {\z}_{c} {\bar\varpi}_{c}+
{\bar\z}_{c}{\varpi}_{c}$, $\star$ is flat space Hodge star, and
\eqn\sd{
\varpi_{r} = {i\over 2}
\left( {\dd}z_1 \wedge    {\dd}\zb_1 +
{\dd}z_2 \wedge {\dd}\zb_2 \right), \,
\varpi_{c} = {\dd}z_1 \wedge {\dd}z_2.}
If $\zeta^{c}=0$ then \sdp\ implies that $F^{0,2}=0$, i.e.
the $A_{\zb_1}, A_{\zb_2}$ define a holomorphic structure on the
bundle ${\CE}_z = {\rm ker}{\CD}^{\dagger}_z$ over $X^{\circ}$.
As we have a unitary
connection, $F^{2,0}=F^{0,2}=0$.

From \exct\ the holomorphic bundle ${\CE}$ extends to a holomorphic bundle
${\CF}$ on  the whole of ${\IR}^4$.
We will now construct a compactification $X$ of $X^{\circ}$ with a
holomorphic bundle ${\tilde\CE}$ over $X$ such that
${\tilde\CE} \vert_{X^{\circ} } \approx {\CE}$,
and whose connection ${\tilde A}$ is a smooth
continuation of the connection $A$ over $X^{\circ}$.
This compactification $X$ projects down to ${\IC}^2$ via a
map $p: X \to {\IC}^2$.
The pull-back $p^*{\CF}$ is a holomorphic bundle over $X$ which differs
from ${\tilde\CE}$.
This difference is localised at the exceptional variety, which is
the preimage $p^{-1}(Z)$ of the set
of freckles.

\newsec{Abelian case in detail}

Let us rotate $\vec \z$ so that
 ${\z}_{c} = 0, \zeta_r = \zeta>0$ and consider the
case $w = 1$.  Then \nakajima\ shows that $J = 0$. Hence,
$I^{\dagger}I = 2v\zeta$ and $[B_{1},B_{2}]=0$.

We can now solve the equations \ferm\ rather explicitly:
\eqn\sli{\pmatrix{\Psi_{1} \cr \Psi_{2} \cr}
= - \pmatrix{ B_{1}^{\dagger} -{\zb}_1 \cr B_{2}^{\dagger} -{\zb}_2 \cr}
{\bx} I {\chi}, }
where
\eqn\bbx{{\bx}^{-1} =  (B_{1}-z_1)(B_{1}^{\dagger}-{\zb}_1)
+ (B_{2}-z_2)(B_{2}^{\dagger}-{\zb}_2)}
and
\eqn\cch{\chi = {1\over{\sqrt{1 + I^{\dagger}{\bx} I}}}.}
Let ${\bf P}(z) =
{\rm Det}{\bf G}^{-1}$. It is a polynomial in $z, {\zb}$ of degree
$v$. Clearly \cch\ implies that:
$$
{\chi}^{2} = {{\bf P}(z) \over {\bf Q}(z)}
$$
where ${\bf Q}(z) = {\bf P}(z) + I^{\dagger} {\widetilde{\bf G}^{-1}} I$
is another degree $v$ polynomial in $z, {\zb}$, ${\widetilde{\bf G}^{-1}}$
being the matrix of minors of ${\bf G}^{-1}$.

The gauge field \gf\ is calculated to be
\eqn\gage{A = (\p - \pb) {\rm log} {\chi},}
and its curvature is
\eqn\crvti{F = \p \pb {\rm log}{\chi}^2.}

The formula \gage\ provides a well-defined one-form on
the complement $X^{\circ}$
in ${\IR}^4$ to the set $Z$ of zeroes of ${\bf P}(z)$.
This is just where $B_1-z_1$ and $B_2-z_2$ fail to be
invertible (and so $\sigma_z$ fails to be injective), that is a ``freckle".
We start with the  study of one such point and then generalize.

\subsec{Charge one instantons.}

To see what happens at such a point let us first look at the case $v=1$.
Then (after shifting ${\zb}_1$ by $B_{1}^{\dagger}$,  {\it etc.})
\eqn\fermi{\Psi_{z} = {1\over{r\sqrt{r^2 + 2\zeta}}}
\pmatrix{{\zb}_{1} \sqrt{2\z}\cr
{\zb}_{2} \sqrt{2\z} \cr r^2 \cr}, \quad \chi = {r\over{\sqrt{r^2 + 2\z}}},}
where $r^2 = \vert z_1 \vert^2 + \vert z_2 \vert^2$.
Thus in this case
$$
{\bf P}(z) = z_1 {\zb}_1 + z_2 {\zb}_2, {\quad} {\bf Q}(z) =
z_1 {\zb}_1 + z_2 {\zb}_2 + 2{\z}
$$
The gauge field is given by (setting $2\zeta = 1$):
\eqn\gge{A = {1\over{2r^2 ( 1 + r^2)}} \left(
z_1 {\dd} {\zb}_1  - {\zb}_1 {\dd} z_1 +
z_2 {\dd} {\zb}_2  - {\zb}_2 {\dd} z_2 \right),}
and
\eqn\crvta{F ={ {\dd}z_1\wedge {\dd}{\zb}_1+
{\dd}z_2\wedge {\dd}{\zb}_2  \over r^2 ( 1+ r^2)}
-
{{1+2r^2}\over{r^4 ( 1+ r^2)^2}}   \sum_{i,j} z_i {\zb}_j {\dd}z_j \wedge
{\dd}{\zb}_i .}

\subsec{Comparison with the non-commutative instanton}

Notice the similarity of the solution \gge\ to the formulae (4.56), (4.61) of
the paper \witsei. It has the same asymptotics both in the $r^2 \to 0$ and
$r^2 \to \infty$   limits. Of course the formulae in \witsei\ were meant to
hold only for slowly varying fields and that is why we don't get precise
agreement. Nevertheless, we conjecture that all our gauge fields
are the transforms of the non-commutative instantons from \neksch\
under the field redefinition described in \witsei.
{}From our analysis below, it follows that one has to modify the topology
of space-time in order to make non-singular
the corresponding gauge fields of the ordinary gauge theory.

\subsec{The first blowup}

To examine \gge\ further let us rewrite $A$ as follows:
$$
A = A_{0} - A_{\infty},
$$
$$A_{0} =
{1\over{2r^2}} \left(
z_1 {\dd} {\zb}_1  - {\zb}_1 {\dd} z_1 +
z_2 {\dd} {\zb}_2  - {\zb}_2 {\dd} z_2 \right),$$
$$A_{\infty} =
{1\over{2(1+r^2)}} \left(
z_1 {\dd} {\zb}_1  - {\zb}_1 {\dd} z_1 +
z_2 {\dd} {\zb}_2  - {\zb}_2 {\dd} z_2 \right).$$
The form $A_{\infty}$ is regular everywhere in ${\IR}^4$.
The form $A_{0}$ has a singularity at $r=0$. Nevertheless, as
we now show, this  becomes a well-defined gauge field on ${\IR}^4$
blown up at one point $z = 0$.

Let us describe the blowup in some details. We start with ${\IC}^2$
with coordinates  $(z_1, z_2)$. The space blown up at the point
$0 = (0,0)$ is simply the space $X$
of pairs $(z,{\ell})$, where
$z \in {\IC}^2$, and ${\ell}$ is a complex line which passes through
$z$ and the point $0$. $X$ projects to ${\IC}^2$ via the map
$p (z, {\ell}) = z$. The fiber over each point $z \neq 0$
consists of a single point while the fiber over the  point $0$
is the space ${\IC\IP}^1$ of complex lines passing through
the point $0$.

In our applications we shall need a coordinatization of the blowup.
The total space of the blowup is a union
$X = {\CU} \cup {\CU}_{0} \cup {\CU}_{\infty}$ of three
coordinate patches. The local coordinates in the patch ${\CU}_0$
are $(t, {\l})$ such that
\eqn\op{z_1 = t, \, z_2 = {\l}t.}
In this patch ${\l}$ parameterises
the complex lines passing through the point $0$, which are not
parallel to the $z_1 = 0$ line.
In the patch ${\CU}_{\infty}$ the coordinates  are
$(s ,{\m})$, such that
\eqn\anp{z_1 = {\m}s, z_2 = s.}
There is also a third patch ${\CU}$, where $(z_1, z_2) \neq 0$.
This projects down to ${\IC}^2$ such that over each point
$(z_1,z_2) \neq 0$ the fiber consists of just one point. The fiber
over the point $(z_1, z_2) =0$ is the projective line
${\IC\IP}^1 = \{ {\l}  \} \cup \infty$. We now show that on this
blown up space our gauge field is well defined.

On ${\CU} \cap {\CU}_{0}$ we may write
\eqn\ggei{A_{0} = {{t {\dd} {\tb} - {\tb} {\dd} t}\over{2\vert t \vert^2}}
+
{{{\l} {\dd} {\bar{\l}} -
{\bar {\l}}{\dd} {\l}}\over{2(1+\vert {\l} \vert^2)}}.}
Define $A_{{\CU}_{0, \infty}}$ as
\eqn\ggeo{\eqalign{
A_{{\CU}_{0}} \quad = & \quad {{{\l} {\dd} {\bar{\l}} -
{\bar {\l}}{\dd} {\l}}\over{2(1+\vert {\l} \vert^2)}},\cr
A_{{\CU}_{\infty}} \quad = & \quad {{{\m} {\dd} {\bar{\m}} -
{\bar {\m}}{\dd} {\m}}\over{2(1+\vert {\m} \vert^2)}}.\cr}}
Now $A_{0}$ is a well-defined one-form on $\CU$.
On the intersections ${\CU} \cap {\CU}_{0}$
the one-forms $A_{0}$ and $A_{{\CU}_{0}}$ are related via
a gauge transformation $$i \,{\dd}\, {\rm arg} t .$$
On the intersection ${\CU}_{0} \cap {\CU}_{\infty}$
the one-forms $A_{{\CU}_0}$ and $A_{{\CU}_{\infty}}$
are related via $$i\, {\dd} \,{\arg} {\l} = - i\, {\dd} \, {\rm arg} {\m}$$
gauge transformations. Finally on ${\CU} \cap {\CU}_{\infty}$
the one-forms $A_{0}$ and $A_{{\CU}_{\infty}}$
are related via the gauge transformation
$$i \,{\dd} \,{\rm arg} s .$$
We have shown therefore that $A_0$ is a well-defined gauge field on
$X$. Observe also that at infinity $A \to 0$ as $o(r^{-3})$, which
yields a finite action. In fact
the gauge field \gge\ has a non-trivial Chern class ${\rm ch}_2$:
\eqn\acti{F \wedge F =  - {2\over r^2 (1 + r^2)^3} {\dd}z_1 \wedge
{\dd}{\zb}_1 \wedge {\dd}z_2 \wedge {\dd}{\zb}_2}
so that
$$
{1\over{4\pi^2}} \int F \wedge F = 1.
$$
Finally, the restriction of $A$ on the exceptional divisor $E$,
defined by the equation $t = 0$ in ${\CU}_0$ and $s = 0$ in ${\CU}_{\infty}$,
has non-trivial first Chern class:
$$
{1\over{2\pi i}} \int_{E} F = - 1.
$$
\subsec{Charge two }

In the case $v > 1$  the formulae \gage, \cch\ are rather intricate.
Nevertheless we show that by a sequence of blowups we are able to
construct a space $X$
on which the formula
\gage\ defines a well-defined gauge field.

For $v=2$ the matrices $(B_1, B_2)$  and the vector $I$ can be brought to the
following normal form by a complexified gauge transformation \scndr\
with $g \in {\rm GL}_2({\IC})$:
\eqn\nrml{B_1 = \pmatrix{ 0 & p_1 \cr 0 & 0}, \quad B_2 = \pmatrix{0 & p_2 \cr
0 & 0 }, \quad I = \pmatrix{0 \cr 1 } }
where the only modulus is $p  = (p_1 : p_2)$ which is a point in ${\IC\IP}^1$.
This data parameterises the torsion free ideal sheaves ${\CI}$ on ${\IC}^2$
which become locally free on the manifold $X$
which is a blowup of ${\IC}^2$ at the point $0$ subsequently
blown up at the point $p$ on the exceptional divisor.

\ndt
{\sl The sheaf and its liberation.}
The ideal ${\CI}_p$ corresponding to $p \in {\IP}^1$ is
spanned by the functions $f(z_1, z_2)$ on ${\IC}^2$  such that:
$$
f \in {\CI}_{p} \Leftrightarrow f(0,0)=0, \quad
p_1 {\p}_1 f \vert_{0} + p_{2} {\p}_2 f \vert_{0} = 0,
$$
i.e. ${\CI}_p = \langle p_1 z_2 - p_2 z_1, z_1^2, z_1z_2,  z_2^2 \rangle$.
This sheaf becomes locally free on the manifold $X$. Indeed, consider
first the manifold $Y$ which is ${\IC}^2$ blown up at the point $(0,0)$.
Suppose $p_1 \neq 0$, hence we may set $p_1 = 1$, $p_2 = p$.
In the chart ${\CU}_0$ where the good coordinates are:
$(z_1, {\l} = z_2/z_1)$ the ideal is spanned by:
$$
z_1 ( p -   {\l}), z_1^2
$$
which is the sheaf ${\CO}(-E)$, where $E = \{ z_1 = 0 \}$ is the exceptional
divisor,
tensored with the ideal sheaf of the point $z_1 = 0, {\l} = p$.
Upon the further blow up $X \to Y$ at the point $\l = p, z_1 = 0$
we get the locally free sheaf, whose sections vanish at the exceptional
variety.

\ndt
{\sl The gauge field construction.} It turns out that the solution of the
equations \mmnts\ up to the
\scndr\ group action does not differ much from \nrml. In fact,
\eqn\nrmli{B_1 = \pmatrix{ 0 & p_1 \cr 0 & 0}, \quad B_2 = \pmatrix{0 & p_2 \cr
0 & 0 }, \quad I = \sqrt{4\zeta} \pmatrix{0 \cr 1 } }
is the solution, provided
$$
\vert p_1 \vert^2 + \vert p_2 \vert^2 = 2\zeta.
$$
Then  formula \gage\ still holds with
\eqn\cchi{\chi^2  =  {{r^2 (r^2 + 2\zeta)  - \vert \eta \vert^2 }\over{
                       (r^2 + 2\zeta)(r^2 + 4\zeta) - \vert \eta \vert^2}},}
where $\eta = z_1 {\bar p}_1 + z_2 {\bar p}_2$. This
expression naively leads to a singular
gauge field when $r^2=0$. (The denominator is nonvanishing for $\zeta>0$.)
We deal with that by blowing up
the point $r = 0$. In this way we can write
$$
r^2 = \vert z_1 \vert^2 ( 1 + \vert \l \vert^2), \quad
\eta = z_1 ( {\bar p}_1 + {\l } {\bar p}_2 )
$$
(in the patch ${\CU}_0$) to  get  a factor
$\vert z_1 \vert^2$ from the numerator of $\chi^2$. This factor is
then removed by a gauge transformation which enters the glueing
function. Then, for $z_1 = 0$,  we find a singularity
at the point ${\l} = p_2 / p_1$ on the exceptional divisor in $Y$
which is removed in a similar way
by the next blowup $X \to Y$.

\subsec{The higher charge case}

The nice feature of the cases $v \leq 2$ is that all information about the
sheaves is encoded in the geometry of the manifolds $X$, $X_p$.
This property may seem to be lost once $v > 2$. Take for example the ideal
${\CI}^{3} = \langle z_1^2, z_1z_2, z_2^2 \rangle$. The quotient
${\IC}[z_1, z_2] /{\CI}^3 = \langle [1], [z_1], [z_2] \rangle$
is three-dimensional.
Similarly the ideal ${\CI}^{4} = \langle z_1^2, z_2^2 \rangle$ produces
a four dimensional quotient ${\IC} [z_1, z_2]/{\CI}^4 = \langle
[1], [z_1], [z_2], [z_1 z_2] \rangle$.
Clearly these ideal sheaves are different. Now consider the first blowup
$X$. Obviously these sheaves then lift to the same sheaf of ideals,
since on $X$: $z_1z_2 = z_1^2 \l \in {\CI}^3$.
Thus we find two different sheaves (for charges 3 and 4) which
lift to the same holomorphic bundle after a single blowup.
The puzzle is what extra data is needed to distinguish them?
Therefore one needs a more refined way of extracting the properties
of the sheaf from the properties of the manifold $X$.
Perhaps the metrics on the blown up space differ
for different charges. We now show that the gauge fields
constructed out of the deformed ADHM data describing these
two ideals are different, so distinguishing them\footnote{$^{!}$}{We were advised by R.~Thomas that at the sheaf level the distinction is captured by the torsion groups $Tor$ one can construct in the course of lifting the sheaf to the blowup}.
\medskip
\ndt
{\sl Charge three gauge fields.}  Consider the ideal ${\CI}$
spanned by
$z_1^2, z_1z_2, z_2^2$. Let us choose the basis
$$
e_1 = 1, \quad e_2 = z_1, \quad e_3 = z_2
$$
in the quotient
$V = {\IC}[z_1, z_2]/{\CI}$. The matrices $B_1, B_2$ act in
$V$ as follows:
$$ B_1 e_1 = e_2, \quad B_1 e_2 = B_1 e_3 = 0,
$$
$$
B_2 e_1 = e_2, \quad B_2 e_2 = B_2 e_3 = 0.
$$
It turns out that a very simple modification makes them a solution
of the equations \mmnts. We find
$$
B_1 = \sqrt{2\z} e_2 e_1^{\dagger},\, \, B_2 = \sqrt{2\z} e_3 e_1^{\dagger},
\,\, I = \sqrt{6\z} e_1,
$$
and consequently
$${\bf G}\sp{-1}= \pmatrix{ r^2 & -z_1\sqrt{2\z}&-z_2\sqrt{2\z} \cr
-{\bar z_1}\sqrt{2\z}&r^2+ 2\z&0 \cr
-{\bar z_2}\sqrt{2\z}&0& r^2+ 2\z} .$$
One finds that
$${\bf G}_{11}={{(r^2+2\z)^2}\over{r^4 (r^2 +2\z)}}$$
and so
\eqn\chthr{{\chi}^2 ={1\over{{1+6\z {\bf G}_{11}}}}=
{{r^4}\over{{(r^2 + 3{\z})^2 + 4 {\z}^2}}}.}

\medskip
\ndt
{\sl Charge four.} Now let us take the ideal ${\CI} = \langle z_1^2, z_2^2
\rangle$. The quotient $V = {\IC}[z_1, z_2]/{\CI}$ is four dimensional
with the basis
$$
e_1 = 1, \, e_2 = z_1, \, e_3 = z_2, \, e_4 = z_1 z_2.
$$
The corresponding solution to the real moment map equations turns out to be
\eqn\rlmfr{B_1 = \sqrt{3\z} e_2 e_1^{\dagger} +\sqrt{\z} e_4 e_3^{\dagger},
\qquad  B_2 = \sqrt{3\z} e_3 e_1^{\dagger} +\sqrt{\z} e_4 e_2^{\dagger},
\qquad I = \sqrt{8\z} e_1.}
Then
$${\bf G}\sp{-1}= \pmatrix{ r^2 & -z_1\sqrt{3\z}&-z_2\sqrt{3\z}&0 \cr
-{\bar z_1}\sqrt{3\z}&r^2+ 3\z&0&-z_2\sqrt{\z} \cr
-{\bar z_2}\sqrt{3\z}&0&r^2+ 3\z&-z_1\sqrt{\z} \cr
0&-{\bar z_2}\sqrt{\z}&-{\bar z_1}\sqrt{\z}& r^2+ 2\z}$$
and
\eqn\chfr{{\chi}^2= {1\over{{1+8\z {\bf G}_{11}}}}=
{{r^4 ( (r^2+2{\z})^2+2{\z}^2) - 12 {\z}^2
\vert z_1 z_2 \vert^2 }\over{((r^2 + 4{\z})^2 + 8{\z}^2)((r^2 + 2{\z})^2+
2{\z}^2) - 12{\z}^2 \vert z_1 z_2 \vert^2}}.}

Clearly this expression is quite different from
\chthr\ and so  the gauge fields do somehow distinguish the
different ideals. Both \chthr\ and \chfr\ require
a single blowup to make the gauge field non-singular.
In the charge three case the gauge field restricted onto the exceptional
divisor looks like $({\p} - {\pb}) {\rm log}
{\chi}_3^{\prime}$ with
${\chi}_3^{\prime} = 1 + \vert {\l} \vert^2$ in the ${\CU}_0$ chart
and $1 + \vert \m \vert^2$ in  ${\CU}_{\infty}$ chart.
In contrast the charge four gauge field on the exceptional divisor has
${\chi}_4^{\prime} = \sqrt{ 1 + \vert {\l} \vert^4 }$ on ${\CU}_0$
and
$\sqrt{1 + \vert {\m} \vert^4}$ on ${\CU}_{\infty}$.
One may say that the exceptional divisor in the charge three case is
more ``rounded'' than the one in the charge four case.

\medskip
\ndt
{\sl Elongated instantons: the general case.} These are
special solutions to the deformed ADHM  equations that describe
$v \geq 1$ points which sit along a complex line.
The ideal ${\CI}$ corresponding to this configuration is (in an
appropriately rotated coordinate system) generated
by $\langle P(z_1), z_2 \rangle$, where $P(z)$ is an arbitrary
 degree $v$ polynomial. In other words the space of elongated
torsion free sheaves of rank one is isomorphic to the
space of degree $v$ polynomials $P$.
If $P(z) = z^{v}$ then we get $v$ points on top of each other.
We shall study this case in some detail a little later after
first presenting the case of general $P(z)$.
Since $z_2 \in {\CI}$ we immediately conclude that $B_2 = 0$.
Then the moment map equation  ${\mu}_r = 2\z$ coincides precisely
with the moment map leading to the Calogero-Moser integrable system \kkk.
In the light of
\nikdui\ its solutions form a part of the phase
space of the Sutherland model. For us the most convenient
presentation is in terms of the dual \dual\ -
rational Ruijsenaars system \ruij.
Consider the polar decomposition $B_1 - z_1 = U^{\dagger} H^{1\over 2}$,
$B_2 = 0$, with $H$ hermitian and positive definite, and $U$
unitary. We may take $B_1$ traceless by appropriately shifting
$z_1$. Then equation \mmnts\ becomes
$$
U^{\dagger} H U - H + I I^{\dagger} = 2{\z}.
$$
This can be solved (as in \relcal) by first diagonalising $H$
\eqn\dh{
H = 2{\z} {\rm diag} \left( r_1^2, \ldots, r_v^2 \right),
}
(with $r_i^2 \geq r_{i-1}^2 +1$) and then solving for $U$ and $I$.
Let ${\CP}(z) = \prod_{i=1}^{v} \left( z - r_i^2 \right)$. Then
with
$$I_i= \sqrt{2{\z}} y_i, \qquad x_i = (U y)_i,$$
we find
\eqn\solu{
U_{ij} = {{x_i {\bar y}_{j}}\over{r_j^2 - r_i^2 + 1}},\qquad
z_1= - {1\over{v}} \sum_{i=1}^{v} {\bar x}_i y_i r_i,
}
where (employing manipulations familiar in Lagrange interpolation)
$$\vert x_i \vert^2 = {{\CP}( r_i^2 - 1) \over - {\CP}^{\prime}(r_i^2) },
\qquad
\vert y_i \vert^2 = {{\CP}( r_i^2 + 1) \over {\CP}^{\prime}(r_i^2) }.
$$
The remaining gauge invariance allows us to make $y_i$ real and non-negative.
The phases of $x_i$ for $r_i^2 > r_{i-1}^2 +1$ are arbitrary, and given by
$$
x_i = \sqrt{{\CP}(r_i^2 - 1) \over - {\CP}^{\prime}(r_i^2)} e^{-i\theta_i}.
$$
The polynomial $P(z)$ which corresponds to the solution \dh, \solu\
is given by
$$
P(z_1) = {\Det} (B_1 - z_1) = \prod_{i} r_{i} e^{i \theta_i}.
$$
Finally, a short calculation shows that for these solutions
\eqn\chalon{{\chi} = \sqrt{{{\CP}\left(-{{\vert z_2 \vert^2)}\over{2\z}} \right)}
\over{{\CP}\left(-{{\vert z_2 \vert^2}\over{2\z}} - 1 \right)}}.}

For the case $P(z)=z^{v}$ the solution  \chalon\ can be made more explicit.
By a change of basis in $V$, the solution to the hyperk\"ahler
moment map equations \mmnts\  acquires a simpler form, viz.
\eqn\slna{B_1 =  \sum_{i=1}^{v-1} \sqrt{2(v-i){\z}}\, e_{i+1} e_{i}^{\dagger},
\qquad  B_{2} = 0, \quad I = \sqrt{2v{\z}}\, e_1.}
Then $P(z) = {\Det} (B_1 - z)=z^{v}$ and
$$
{\bf G}^{-1} =  r^2 +
 \sum_{i=1}^{v} \left( 2(v-i){\z} e_{i+1} e_{i+1}^{\dagger}
-  \sqrt{2(v-i){\z}} \left( z_1 e_{i} e_{i+1}^{\dagger}
+ {\zb}_1 e_{i+1} e_{i}^{\dagger} \right)\right).
$$
Observe that ${\bf G}^{-1}$ is a tridiagonal matrix. Now
in order to find $\chi$ we again need $e_{1}^{\dagger} {\bf G} e_{1}$.
This is easily done, the tridiagonal nature of ${\bf G}^{-1}$ reducing
the problem to one of a three term recursion. Suppose $u_k$ satisfies
\eqn\recur{-  \sqrt{2{\z}(k+1)}\, \zb_1 u_{k+1}+
 \left(r^2 + 2{\z} \left( k+1\right)\right)\, u_{k}
-\sqrt{2{\z}k}\, {z}_1 u_{k-1}=0. } Then $${\bf G}^{-1}
(u_{v-1},u_{v-2},\ldots u_0)\sp{t}=(r^2 u_{v-1}-{z}_1
\sqrt{2{\z}(v-1)}u_{v-2},0,\ldots 0)\sp{t}$$ and consequently
$${\bf G}_{11}= {u_{v-1}/( r^2 u_{v-1}-{z}_1
\sqrt{2{\z}(v-1)}u_{v-2})}.$$ The normalisation of the $u_k$'s is
irrelevant. Now the substitution $$u_k= {{{z}_1\sp{k}}
\over{\sqrt{(2 {\z})\sp{k} k!}}} w_{k}$$ simplifies \recur\ to
give \eqn\charl{ x w_{k+1}- (x + y +1 +k) w_k +k w_{k-1}=0,} where
we have set $\vert z_1 \vert^2 = 2{\z} x, \vert z_2 \vert^2 =
2{\z} y$. Together with the normalization $w_0=1$, we recognize
the recursion for the Charlier polynomials $w_k = C_k (-1-y;
x)=\phantom{.}_2F_0( {-k,-1-y \atop -};{-1\over x})$. These may
also be expressed in terms of the Laguerre polynomials as
$(-x)\sp{k} C_k(a;x)/{k!}= L_k\sp{(a-k)}(x)$. They have the
generating function $$ e\sp{t}\left(1-{t\over
x}\right)\sp{a}=\sum_{k=0}\sp\infty {C_k(a;x)\over k!}t\sp{k}. $$
In terms of $w_k$, we find $${\bf G}_{11}= {1\over
2{\z}}{w_{v-1}\over {r^2\over 2{\z}}w_{v-1}-(v-1)w_{v-2}},\qquad
1+I^{\dagger} {\bf G}I= x {w_{v}\over{r^2\over
2{\z}}w_{v-1}-(v-1)w_{v-2}}.$$ Differentiation of the generating
function shows that $$ (x+y) C_k(-1-y;x)-k C_{k-1}(-1-y;x)=x
C_{k+1}(-y;x)$$ from which it follows that \eqn\chaltw{{\chi} =
\sqrt{C_v(-y;x) \over{ C_v(-y-1;x)}}.} One may show that the
polynomial ${\CP}$ corresponding to this special solution of the
ADHM equations is simply $$ {\CP}(z) = C_v(z;x). $$

\newsec{Non-abelian charge one freckled instantons}

We now proceed with the investigations of the non-abelian case,
considering the example of charge one instantons. The deformed
ADHM construction in the case $v=1$ gives the space
${\overline{M}}_{1,w} \approx {\IR}^4 \times T^{*}{\IC\IP}^{w-1}$.
The first factor is the space of pairs $(B_{1},B_{2})$ which
parameterize the center of the instanton. The second factor is
responsible for its size and orientation.

Specifically, the  second factor emerges  as a quotient
of the space of pairs
$(I, J)$, $I \in W^{*}, J \in W$, such that
\eqn\mmnts{IJ = 0, \quad I\, I^{\dagger} - J^{\dagger}\, J = 2\z > 0}
by the action of the group $U(1)$
\eqn\scndg{
(I,J) \mapsto (Ie^{i\t}, J e^{-i\t}).}
Let us introduce two projectors
\eqn\abcd{P_1 = I^{\dagger} I, \quad P_2 = JJ^{\dagger},}
and two numbers
\eqn\aib{{\rho_1}^2 = I I^{\dagger}, \quad {\rho_2}^2 = J^{\dagger}J.}
Then ${\rho_1^2} - {\rho_2^2} = 2{\z}$.
In particular $\rho_1 > \rho_2 \geq 0$, and if ${\rho}_2 > 0$ we can write
$$
I^{\dagger} = \rho_1 e_{1}, J = \rho_2 e_{2},
$$
where $e_{1}, e_{2}$ form an orthonormal pair of vectors in $W$.
We shall distinguish between the $\rho_2=0$ and
$\rho_2 >0$ cases in what follows.

Let us proceed with the ADHM construction. Without any loss
of generality we may assume that $B_{1} = B_{2} = 0$, by shifting
$z_1, z_2$.
The vector $\Psi_z : W \to {\IC}^2 \oplus W$
is found to be
\eqn\vctrps{\eqalign{\Psi_z =  {1\over r^2}
\pmatrix{{\zb}_1 I - z_2 J^{\dagger} \cr
{\zb}_2 I + z_1 J^{\dagger} \cr r^2 \cr} {\chi},\quad &
{\chi} = {r \over \sqrt{ r^2 + P_1 + P_2} } =  \cr = \,
1 \, + & \, {P_1 \over \rho_1^2} \left( {r\over \sqrt{r^2 + \rho_1^2}} - 1 \right)
+ {P_2 \over \rho_2^2} \left( {r\over \sqrt{r^2 +\rho_2^2}}- 1 \right) ,\cr}}
where in the process of solving for $\Psi_{z}$ we used the gauge
$\chi^{\dagger} = \chi$.

Notice that in order to write the explicit formula \vctrps\
for the vector
$\Psi_z$ we had to make a choice of the vectors $I,J$ in the
orbit \scndg. When working on flat ${\IR}^4$ this choice
can be made globally, i.e. in a $z$ independent way.
If we are to replace ${\IR}^4$ by a manifold
$X$ over which  non-trivial line bundles exist,
then this choice may well become a subtle matter,
i.e. the solution to the equations \mmnts\ may depend on $z$
while staying in the orbit of the gauge group
\scndr. In other words $\theta$ may depend on $z$.
In a moment we shall see
that this indeed  happens.

Using the relations $I\chi = {r\over \sqrt{r^2 + \rho_1^2}} I$,
and $J^{\dagger}{\chi} = {r\over \sqrt{r^2 + \rho_2^2}} J^{\dagger}$,
we may write
\eqn\vctrpsi{\Psi = \pmatrix{{{\zb}_1 \over {r \sqrt{r^2 + \rho_1^2}}} I -
{z_2 \over{r \sqrt{r^2 + \rho_2^2}}} J^{\dagger} \cr
{{\zb}_2 \over{r \sqrt{r^2 + \rho_1^2}}} I +
{z_1 \over{r \sqrt{r^2 + \rho_2^2}}} J^{\dagger} \cr
\chi \cr}.}
This expression is well-defined for $r \neq 0$. Moreover $\chi$ is
well-defined everywhere, while $\Psi_{1}, \Psi_{2}$ have singularities
at $r=0$.
Let us perform a sigma process
at  $(z_1, z_2) = 0$. Introduce the coordinates
$(t,\l)$ and $(s, \m)$ by the formulae \op, \anp.

\ndt {\sl The locally free sheaves: $\rho_2 > 0$.} In this case we may
write
\eqn\splt{\chi = \chi^{\perp} + \chi^{\Vert}, \qquad \chi^{\Vert} =
{r\over{\sqrt{r^2 + {\rho_1^2}}}}
e_{1} e_{1}^{\dagger} +
{r\over{\sqrt{r^2 + {\rho_2^2}}}}
e_{2} e_{2}^{\dagger}.}
The component ${\chi}^{\perp} = (1 - e_{1}e_{1}^{\dagger} -
e_{2}e_{2}^{\dagger})\chi$ decouples.
In this sense, it is sufficient to study the case $w=2$ only.
We are free to perform a
gauge transformation on  $\chi^{\Vert}$ and ${\vf}$ not affecting
the $\chi^{\perp}$ part.

\ndt
In the patch ${\CU}_{0}$ we may write
\eqn\opp{\vf_{0} = {1\over{\sqrt{1+\vert \l \vert^2}}}
\pmatrix{
{I \over \sqrt{r^2 + \rho_1^2}} - {{\l} J^{\dagger} \over
\sqrt{r^2 + \rho_2^2}} \cr
{{\bar\l} I \over \sqrt{r^2 + \rho_1^2}} +
 { J^{\dagger} \over \sqrt{r^2 + \rho_2^2}}\cr},}
while in the patch ${\CU}_{\infty}$ we similarly have
\eqn\anpp{\vf_{\infty} = {1\over{\sqrt{1+\vert \m \vert^2}}}
\pmatrix{
{{\bar\m} I \over \sqrt{r^2 + \rho_1^2}} - { J^{\dagger}
\over \sqrt{r^2 + \rho_2^2}} \cr
{ I \over \sqrt{r^2 + \rho_1^2}} +
 {{\m} J^{\dagger} \over \sqrt{r^2 + \rho_2^2}}\cr }.}
The gluing across the intersection ${\CU}_{0} \cap {\CU}$ is
achieved with the help of a $U(w)$ gauge transformation which acts
on the vectors $e_{1}, e_{2}$ only, leaving $\chi^{\perp},
\chi^{\Vert}$ unchanged, \eqn\gone{g_{{\CU}{\CU}_{0}}
\pmatrix{e_{1}\cr e_{2}\cr} = \pmatrix{{t\over{\vert t\vert}}
e_{1}\cr  {{\tb} \over{\vert t\vert}} e_{2}\cr},} so that $\vf =
\vf_{0} g_{{\CU}{\CU}_{0}}^{\dagger}$. Analogously,
\eqn\gtwo{g_{{\CU}{\CU}_{\infty}} \pmatrix{e_{1}\cr e_{2}\cr} =
\pmatrix{{{s} \over{\vert s\vert}} e_{1} \cr  {{\sb} \over{\vert
s\vert}} e_{2}\cr}, \qquad g_{{\CU}{\CU}_{0}} =
g_{{\CU}{\CU}_{\infty}}^{-1}.} Finally,
$g_{{\CU}_{0}{\CU}_{\infty}} = g_{{\CU}{\CU}_{\infty}}^{2}$.

In the patch $\CU$ the  gauge field is given by
$$
A =
e_1 e_1^{\dagger} \left( {\pb} - {\p} \right)
{\rm log}{r \over \sqrt{r^2 + \rho_1^2}}
-
e_2 e_2^{\dagger} \left( {\pb} - {\p} \right)
{\rm log}{r \over \sqrt{r^2 + \rho_2^2}}
$$
\eqn\ggef{
+{\rho_1 \rho_2 \over \sqrt{\left( r^2 + \rho_1^2\right)
\left( r^2 + \rho_2^2\right)}}
\left(
e_2 e_1^{\dagger} {{{\zb}_1 {\dd} {\zb}_2 - {\zb}_2 {\dd} {\zb}_1 }\over
{r^2}} +
e_1 e_2^{\dagger} {{{z}_2 {\dd} z_1 - z_1 {\dd} z_2 }\over
{r^2}} \right)}
and its field strength
\eqn\fstr{\eqalign{F\quad  = \quad &\sum_{i,j}
e_i e_j^{\dagger} F_{ij}, \cr
& F_{11} = {\p\pb} {\rm log} {r^2 \over r^2 +\rho_1^2} - {{\rho_1^2
\rho_2^2}\over{\left( r^2 +
\rho_1^2\right)\left( r^2 + \rho_2^2\right)}} {\p\pb} {\rm log} r^2, \cr
& F_{22} = - {\p\pb} {\rm log} {r^2 \over r^2 +\rho_2^2} + {{\rho_1^2
\rho_2^2}\over{\left( r^2 +
\rho_1^2 \right)\left( r^2 + \rho_2^2 \right)}} {\p\pb} {\rm log} r^2, \cr
& F_{12} = - 2{{\left( 2r^2 + {\rho}_1^2 + {\rho_2^2}\right)
{\rho_1 \rho_2}}\over{r^4 (r^2+{\rho_1^2})^{3/2}(r^2+{\rho}_2^2)^{3/2}}}  \left( z_1 {\dd}z_2 -
z_2 {\dd}z_1 \right) \wedge {\pb} r^2,\cr
& F_{21} = - {\overline{F_{12}}}. \cr}}

In the patch ${\CU}_{0}$ the expression for the gauge field is modified
to
$$
A =
e_1 e_1^{\dagger} \left( {\pb} - {\p} \right)
{\rm log}{1 + \vert \l \vert^2 \over \sqrt{r^2 + \rho_1^2}}
-
e_2 e_2^{\dagger} \left( {\pb} - {\p} \right)
{\rm log}{1 + \vert \l \vert^2 \over \sqrt{r^2 + \rho_2^2}}
$$
\eqn\ggef{
+{\rho_1 \rho_2 \over \sqrt{\left( r^2 + \rho_1^2
\right) \left( r^2 + \rho_2^2
\right)}}
\left(
e_2 e_1^{\dagger} {{ {\dd} {\bar\l}  }\over
{1 + \vert \l \vert^2}} +
e_1 e_2^{\dagger} {{  -  {\dd} \l }\over
{1 + \vert \l \vert^2}} \right).}

\ndt {\sl The freckles: $\rho_2 = 0$.} This value of the parameter
$\rho_2$ corresponds to the torsion free
sheaves which are not locally free. For $\rho_2 =0$ the vector $J$ vanishes,
${\rho}_1^2 = 2\z$, $I = \sqrt{2\z} e^{\dagger}$, $e \in W$
and the vector $\Psi_z$ simplifies to (removing the
$\chi^{\perp}$ part):
\eqn\nfr{\Psi_z =
{1\over{r \, \sqrt{r^2 + 2\z}}}
\pmatrix{\sqrt{2\z} {{\zb}_1} e^{\dagger} \cr
\sqrt{2\z} {{\zb}_2} e^{\dagger} \cr
{r^2} e e^{\dagger} \cr}}
We easily recognize the vector $\Psi_z$ from the abelian section.
So in this case the torsion free sheaf of rank $w$ splits as
a direct sum of the trivial holomorphic bundle of rank $w-1$ and
a standard charge one torsion free sheaf of rank $1$ which lifts
to a line bundle on the blowup.

\newsec{Discussion}
Our paper has been concerned with the deformed ADHM equations. These equations
may be viewed as giving instantons on a noncommutative space-time. Equally,
and this is the focus of our paper, they may be interpreted as
gauge fields on
an ordinary commutative space-time manifold ${\IC}^2$
or ${\IC\IP}^2$ blown up at a finite
number of points. The deformation singles out a particular complex structure.
In terms of this complex structure, the deformed ADHM construction yields
a holomorphic bundle ${\CE}_z = {\rm ker}{\CD}^{\dagger}_z$
outside of a set of points where $\sigma_z$ fails to be surjective.
The constraint $F^{0,2}=0$ yielding this holomorphic structure may be viewed
as a restriction for supersymmetry.
In addition ordinary instantons obey the equation $F^{1,1 +} = 0$,
which is usually viewed as fixing the ``non-compact''
part of the complexified gauge group. Our solutions apparently
obey another equation ${\CZ}(F^{1,1}) = 0$
which also serves as a gauge fixing condition.
At the points for which
$\sigma_z$ fails to be surjective (elsewhere called ``freckles")
the ADHM gauge fields look singular, and we have shown that by suitably
blowing up such points the gauge fields may be extended in a regular manner.
Our construction is consistent with the work of Seiberg and Witten who
show that (to any finite order) there is a mapping from ordinary gauge
fields to non-commutative gauge fields that respects gauge equivalence.
Presumably the equation ${\hat F}^{1,1 +} = 0$ is mapped into our
equation ${\CZ}(F^{1,1})$ which we admittedly weren't able to identify in
full generality (perhaps the results of \mmms\ may help to solve this
problem).

Our blowups will only be seen at short wavelengths and regulate
the divergences encountered by Seiberg and Witten (cf. Section 4
of \witsei).  We believe the modifications to the topology of
space-time we have described are necessary in order to make the
corresponding gauge fields of the ordinary gauge theory
non-singular. For large instanton number we interpret our results
as producing space-time foam.

Although our study has focused on the $U(1)$ situation,
we have shown how one may extend to the non-abelian situation.
Several of the $U(1)$ constructs reappeared in that case.
As well as being rather concrete, the $U(1)$ situation has revealed a rather
rich structure.
Our low order computations show how the gauge fields can distinguish
between two different sheaves that lift to the same holomorphic bundle
after blowup.
We have also described a general class of instantons that
we called {\it elongated}.
We  were able to associate to this class of solution precisely the moment
map equation of the Calogero-Moser integrable system and used the
machinery of integrable systems to describe this case in some detail. This
appearance of the Calogero-Moser system is somewhat different to that
of Wilson \wilson. This appearance of integrable systems here, in
Seiberg-Witten theory more generally, and in the various brane descriptions
of these same phenomena, still awaits a complete explanation.

\subsec{Notes added in a year} \lref\fk{K.~Furuuchi,
hep-th/9912047, hep-th/0005199, hep-th/0010006}
\lref\nekrev{N.~Nekrasov, ``Noncommutative instantons revisited'',
hep-th/0010017}

After this paper was posted in the archives, a few papers appeared
which addressed the issue of the non-singularity of the
noncommutative instantons more thoroughly. It was shown
\fk\nekrev\ that there is indeed a memory of the blowup of the
commutative space in the noncommutative description (through the
appearence of the shift operators $S$ and $S^{\dagger}$ \nekrev).
However, the
 very noncommutative space over which the instantons are defined,
 is not altered in any way. In this way the noncommutative
 description is simpler, though a physical mechanism for the 
topology change in the commutative description we have encountered 
may be forthcoming.

On the other hand, there was some progress in the search for the
equations ${\CZ}(F^{1,1})= 0$, replacing the ordinary $F^{1,1 \,
+} = 0$. We found that the charge one $U(1)$ gauge field given by
\gge\ is in fact anti-self-dual in Burns\burns\ metric on the blowup of
${\bf R}^4$. The higher charge case however remains open.

\centerline{\bf Acknowledgements}

We are indebted to A.~Rosly and E.F.~Corrigan for numerous helpful
discussions, and to M.~F.~Atiyah, D.~Calderbank, V.~Fock,
M.~Kontsevich, A.~Losev, D.~Orlov, A.~Vainshtein, A.~Schwarz,
S.~Shatashvili, and R. ~Thomas for useful advice.

H.~W.~B. thanks the Royal Society for a grant with the FSU that
enabled this work to begin. Research of N.~N is supported by a
Dicke Fellowship from Princeton University, partly by NSF under
grant PHY94-07194, partly by RFFI under grant 98-01-00327 and by
grant 96-15-96455 for scientific schools. N.~N. also thanks Erwin
Schr\"odinger Institute in Vienna, LPTHE at Universit\'e Paris VI,
ITP, UC Santa Barbara, Royal Society and Universities of Edinburgh
and Heriot-Watt for their support and hospitality during the
course of this work.

\listrefs
\bye